\documentclass[12pt]{article}
\usepackage{epsfig}
\usepackage[hang,bf,small]{caption}

\DeclareGraphicsExtensions{.eps.gz,.eps,.ps,.ps.gz}
\parskip=\itemsep               
\newcommand{\beq}{\begin{equation}}
\newcommand{\eeq}{\end{equation}}
\newcommand{\beqar}[1]{\begin{eqnarray}\label{#1}}
\newcommand{\eeqar}{\end{eqnarray}}

\newcommand{\as}{\alpha_S}

\def\eq#1{{Eq.~(\ref{#1})}}  
\def\npb#1#2#3{    {\it Nucl. Phys. }{\bf B#1} (19#2) #3}
\def\plb#1#2#3{    {\it Phys. Lett. }{\bf B#1} (19#2) #3}
\def\prd#1#2#3{    {\it Phys. Rev. }{\bf D#1} (19#2) #3}

\def\zpc#1#2#3{    {\it Z. Phys. }{\bf C#1} (19#2) #3}

\newcommand{\di}{\partial}
\newcommand{\baralpha}{\bar{\alpha}_S}

\relax
\begin{document}
\title{
\hfill DESY 99-108 \\
\hfill TAUP  2592 - 99\\
\hfill {\tt hep-ph/9908317}\\[0.5cm]
{\Huge  \bf  Solution to the evolution equation }\\
{\Huge \bf for  high parton density QCD}}

\author{{\large \bf  E. ~L e v i n \thanks{E-mail :
\,\,leving@post.tau.ac.il }~ 
$\mathbf{{}^{a),b)}}$} \quad \,\,\,{ \large \bf and \,\, K. ~T u
c h i n \thanks{E-mail : \,\,tuchin@post.tau.ac.il }~ $\mathbf{
{}^{a)}}$}\\[6mm]
{\it ${}^{a)}$ HEP Department}\\
{\it  School of Physics and Astronomy}\\
{\it Raymond and Beverly Sackler Faculty of Exact Science}\\
{\it Tel Aviv University, Tel Aviv, 69978, ISRAEL}\\[1.5ex]
{\it ${}^{b)}$ DESY Theory Group}\\
{\it 22603, Hamburg, GERMANY}}

 \date{}
 \maketitle
\thispagestyle{empty}

\vspace{1cm}
\begin{abstract}
In this paper a solution  is given to the nonlinear  equation
which describes the evolution of the parton cascade in the case of  the  
high
parton density. The related physics is discussed as well as some
applications to heavy ion-ion collisions. 
\end{abstract}

\newpage
\section{Introduction}
\setcounter{equation}{0}
During the past two decades one of the challenging problem of QCD has
 been to understand
theoretically and to observe experimentally a new nonperturbative regime -
high parton density QCD. It has been argued \cite{GLR} in perturbative QCD
(pQCD) 
 that at low $x$ ( high energies )  the
density of parton increases in deep inelastic scattering ( DIS ) and
reaches so
high value that partons 
become densely populated in a hadron. Certainly,  high  density
system of partons  cannot be treated perturbatively but pQCD leads to a
new scale ( mean transverse momentum of partons )  for high parton density
QCD (hdQCD ) as well as to a
hypothesis of parton saturation \cite{GLR}. Intensive theoretical studies
led to deeper understanding of physics of this system as well as to
development of both perturbative 
\cite{PQCDR} and nonperturbative  \cite {NPQCDR} methods for such a
system. However, only recently the first indications on parton saturation
have appeared in HERA data on $Q^2$ - behaviour of the $F_2$-slope
 and on energy behaviour of inclusive diffractive dissociation
in DIS \cite{ABCA}. These data as well as their theoretical interpretation
\cite{THEORINT} mark a new stage of our approach which needs more
quantitive methods, than it was before, to make a reliable prediction for
the experimental observations.

This quantitive approach includes at least two steps:
\begin{enumerate}
\item\,\,\, a  derivation of
the
equation which will be valid in the full kinematic region;
\item\,\,\,
finding a
solution to this equation. 
\end{enumerate}
There are two versions of the nonlinear evolution equation at
$x\,\,\rightarrow\,\,0$. The first one  was suggested as an obvious
generalization of pQCD approach \cite{AGL} while the second has been
obtained from semiclassical gluon field approach \cite{KOVNER}. Both of
them describe correctly the DGLAP \cite{DGLAP} evolution in the limit of
low parton density as well as the GLR nonlinear evolution equation
\cite{GLR} in the region of intermediate parton densities. They give the
same limit of parton density saturation at very low values of $x$ but they
are different in particular description how system reaches this
saturation. Yu. Kovchegov has recently  proved \cite{KOV}  that the GLR
equation
itself (in the form of Eq.(2.18) in Ref. \cite{GLR} ) is able to describe
the QCD evolution in the region of high parton density or, in other
words, at very low values of $x$.   His arguments, which we review briefly
in the next section, are based on Mueller's idea \cite{MUDIPOLE} that
colour dipoles rather than quarks and gluons, are correct degrees of
freedom at high energies ( low $x$ ) in QCD. It was also shown that the
solution to the equation, suggested in Ref.\cite{AGL}, coincides with the
solution to the GLR equation. 

The goal of this paper is to solve the GLR equation, written in the form
suggested in Ref. \cite{KOV}, in the full kinematic region including
$x\,\rightarrow \,0$. The paper is organized as follows: In the next
section we briefly discuss the nonlinear evolution equation for the high
parton density QCD and we formulate our approach for searching a solution
to this equation. In section 3 we introduce the new scale $Q_0(x)$ , which
appears in hdQCD as a solution to the nonlinear equation, and we find the
general solution to the equation for parton with $Q^2\,\,>\,\,Q^2_0
(x)$,
based on the approach developed in Ref.\cite{LALE}. Section 4 is devoted
to a solution of the hdQCD equations in the region of very low $x$. We
show that the parton density reaches the saturation limit and we find the
explicit analytic expression describing  how the system approaches the
saturation
limit. Our summary and conclusions are given in section 5.
\section{Nonlinear equation for high density parton system}
\setcounter{equation}{0}
\subsection{Kinematics, notations  and definitions.}
In this subsection we introduce all notations and definitions as well as
some
kinematic relations that we will need in what follows.  In general, we
try to use the same notation as in Ref.\cite{KOV}. We would like to
clarify the relations between different notations and definitions that have
been used in this area of activity. This task is rather complicated since
 different notations have been used  for the same physics observables,
mostly
due to different theoretical background of authors.  In this paper we
explore the main physical idea of Ref. \cite{MUDIPOLE}, namely, that
correct degrees of freedom in QCD at low $x$ ( high energy ) are colour
dipoles rather than quarks and gluons that explicitly written in the QCD
Lagrangian.  This statement means that a QCD interactions at high energy 
do not change the 
size and the  energy of a colour dipole. Therefore, the majority of our
variables and observables is related to distributions and interactions of
the colour dipoles in a hadron.

\begin{enumerate}
\item\,\,\,$\mathbf{ x_{i\,k} = | \vec{x}_i \,-\,\vec{x}_k |}$ is the
 size of the dipole which consists of a quark $``i"$ at 
$\mathbf{\vec{x}_i}$ and an antiquark $``k"$ at $ \mathbf{\vec{x}_k}$,
or, in other words, it is the transverse separation between quark and
antiquark in a colour dipole;

\item\,\,\,$r\,\,=\,\,\ln
\frac{\mathbf{x^2_0}}{\mathbf{x^2}}\,\,=\,\,\ln\frac{Q^2}{Q^2_0}$, where
$Q^2$ ( $Q^2_0$) is the  transverse momentum of quark in the dipole with
size $\mathbf{ x }$  ( $\mathbf{ x_{0} }$ ) respectively;

\item\,\,\, $x$ is the Bjorken variable for a dipole , $ x = Q^2/W^2$,
where $W$ is the dipole energy;

\item\,\,\, $y = \ln(x_0/x)$, where $x_0$ is  defined for
us the region of low $x$. We consider that for $x \,<\,x_0$ all the
typical features of low $x$ physics should be seen. Practically,
$x_0\,\approx\,10^{-2}$;

\item\,\,\, $xG(x,Q^2)$ is the gluon structure function for a nucleon;

\item\,\,\, $xG_A(x,Q^2)$ is the gluon structure function for a nucleus
with A nucleons;

\item\,\,\, $b_t$ is the impact parameter for the dipole scattering;

\item\,\,\, The main physical observable, which we are going to discuss in
this paper, is the density of dipoles ( $N({ \mathbf{x}},b_t,y )$) with
the
size $\mathbf{x}$ and energy $x$ ( $y$ ) at the impact parameter $b_t$;  

\item\,\,\, In the parton approach ( parton $\equiv $ dipole )   the
dipole
density  $N({\mathbf{x}},b_t,y )  $ is related to the dipole scattering
amplitude if we assume that this amplitude is dominantly imaginary.
Therefore,  $ N( {\mathbf{x}},b_t,y )$ can be normalized using the
unitarity
constrains:
\beq \label{UNITARITY}
 2\,\,N( {\mathbf{x}},b_t,y )\,\,\,=\,\,\,N^2 ( {\mathbf{x}},b_t,y
)\,\,\,+\,\,G^{in}( {\mathbf{x}},b_t,y )\,\,,
\eeq
where $G^{in}$ is the contribution of the inelastic processes to the
scattering of a dipole;
\item\,\,\, From \eq{UNITARITY} one can see that $N( \mathbf{x},b_t,y
)\,\,\leq\,\,1$. Therefore, saturation of the parton density means that
\beq \label{SATURATION}
N({ \mathbf{x}},b_t,y )\,\,\,\longrightarrow\,\,\,
1\,\,\,at\,\,\,x\,\,\,\longrightarrow\,\,\,0\,\,;
\eeq
\item\,\,\,Parameter $\kappa $ which was introduces in many papers
\cite{PQCDR} ( or parameter $W$ of Ref. \cite{GLR} ) is equal to
\beq \label{KAPPA}
\kappa( {\mathbf{x}},y )\,\,\,\equiv W ({ \mathbf{x}},y )\,\,\,=\,\,\, N(
{\mathbf{x}},b_t = 0,y );
\eeq
\item\,\,\,The relation between parton density $N( {\mathbf{x}},b_t,y )$
 and gluon structure function $xG_A (x,\frac{1}{{\mathbf{x^2}}})$
 is given by \cite{AGL} \cite{KOV}
\beq \label{NVSG}
N( {\mathbf{x}},b_t=0,y )\,\,\,\equiv\,\,\,\kappa( {\mathbf{x}},y
)\,\,\,=\,\,\,\frac{\as
\,\pi^2\,{\mathbf{x^2}}}{2\,N_c\,\pi\,R^2_A}\,\times\,xG_A
(x,\frac{1}{{\mathbf{x^2}}}),
\eeq
which holds only for small $N $. We would like to stress that
we do not
need to know the gluon structure function for $N\,\approx\,1$ since $N$
itself has a clear meaning as the scattering amplitude for a dipole. It
should be also stressed that \eq{NVSG} shows that $N( {\mathbf{x}},b_t=0,y
)$ is not a parton ( colour dipole )  density
 $\rho = xG_A(x,\frac{1}{{\mathbf{x^2}}})/\pi R^2_A$ but it is the
scattering  amplitude which is equal to $\sigma_{dipole}\,\times\,\rho $;

\item\,\,\,Widely used function $\phi (k^2,y)$ (see Eq.(2.18) of Ref.
\cite{GLR} for example ) is equal to momentum image of $N$, namely,
\beq \label{NVSPHI}
 \int d^2 b_t \,\,N( {\mathbf{x}},b_t,y )\,\,=\,\,\int \,d^2 k
\,\{\,1\,-\,e^{i
\vec{k}\cdot \vec{b_t}}\,\}\,\,\phi(k^2,y)\,\,;
\eeq
\item\,\,\, $\bar \as $ is used for $\bar \as = \as N_c/\pi$;

\item\,\,\, $R_A$ is the radius of a nucleus in the Gaussian
parameterization of the nucleon density. In this parameterization
the profile function in $b_t$ - representation looks as
\beq \label{PROFFUN}
S_A (b_t )\,\,\,=\,\,\,\frac{1}{\pi R^2_A}\,\,e^{ -
\frac{b^2_t}{R^2_A}}\,\,.
\eeq
$R^2_A
(Gaussian)\,\,=\,\,\frac{2}{5} R^2_A (Wood-Saxon )$, where $R^2( Wood -
Saxon ) \,=\,r^2_0 A^{\frac{2}{3}}$ with $r_0 = 1.3\,fm$ (see
Ref.\cite{AGL} for more details );

\item\,\,\,All physical obsevables, related to nucleus, are denoted with
subscript $A$ while the nucleon observables will be marked by subscript
$N$. Both of these subscripts could be missed if the meaning of the
physical quantity is obvious; 

\item\,\,\, As has been shown \cite{GLR} \cite{PQCDR}, the interaction
between partons leads to a new scale $Q^2_{cr}(x,b_t)=
1/{\mathbf{x^2_{cr}}}$ for hdQCD. The line $r =
\ln(1/{\mathbf{x^2}})\,=\,\ln Q^2_{cr} (x,b_t)$ is called {\it critical
line} ( see Fig.1 ) and all physical quantities defined or calculated on
the critical line are denoted with the subscript {\it cr }. For example,
we denote a new scale, which has meaning of the average parton transverse
momentum in hdQCD parton cascade, by $Q^2_{cr}(x,b_t)$ while in other
papers it has a different notation: $Q^2_{cr}(x,b_t) \,\equiv \,Q^2_0(x)
\cite{GLR}\cite{PQCDR}, \,\equiv\,Q^2_s(x) \cite{MU99}
\,\equiv\,\mu^2(x) \cite{NPQCDR} \cite{KOVNER}$;

\item\,\,\,All physical quantities defined to the right of the critical
line
( for $r \,> \,r_{cr}\,=\,\ln Q^2_{cr}(x,b_t) $) will carry subscript 
$\rightarrow$;

\item\,\,\,All physical quantities defined to the left  of the critical
line
( for $r \,<\, r_{cr}\,=\,\ln Q^2_{cr}(x,b_t)$ ) will carry subscript $   
\leftarrow $.

\end{enumerate} 
\begin{figure}[htbp]
\begin{center}
  \epsfig{file=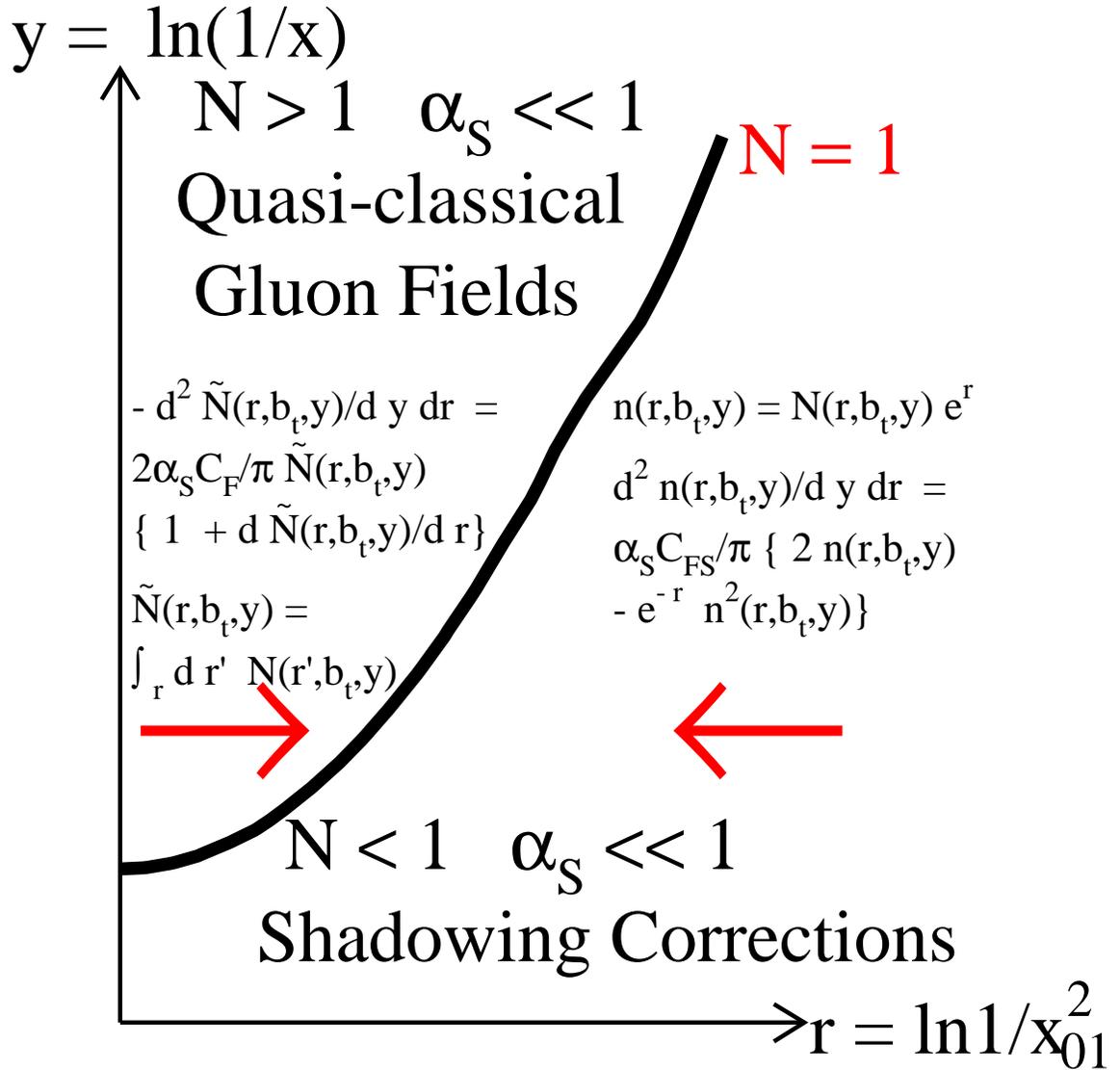,width=170mm}
  \caption[]{\it Different regions for the solution to the hdQCD
evolution equation in the kinematic plot of DIS. The equation
$N({\mathbf{x}},b_t,y) \,=\,1$ gives the critical line. }
 \end{center}
\label{fig1}
\end{figure}

\subsection{ The nonlinear equation for hdQCD evolution }

Yu. Kovchegov in his proof \cite{KOV}, that the GLR  equation is able  
to describe the whole kinematic region including very low values of $x$,
uses heavily two principle ideas suggested by A. Mueller \cite{MUDIPOLE}:
\begin{itemize}
\item\,\,\,The  QCD interaction at high energy does not change the 
transverse separation between quark and antiquark ( the
colour dipole size ), and, therefore, colour dipoles can be considered as
correct degrees of freedom at high energies which diagonalize the strong
interaction matrix;
\item\,\,\,The process of interaction of a dipole with the target has
two clear  stages:
\begin{enumerate}
\item\,\,\,Decay of the dipole into two dipoles, which is described
by 
\beq \label{PSIDIPOLE}
| \Psi(\,
{\mathbf{x_{01}}}\,\rightarrow\,{\mathbf{x_{02}}}\,\,+\,\,{\mathbf{x_{12}}}\,)
|^2\,\,\,=\,\,\,\frac{{\mathbf{x^2_{01}}}}{{\mathbf{x^2_{02}}}\,
{\mathbf{x^2_{12}}}}\,\,;
\eeq
\item\,\,\, Interaction of each  dipole with the target with
amplitude $N({\mathbf{x}},b_t,y)$.

\end{enumerate}

\end{itemize}

The equation  is pictured in Fig.2 and it has the following analytic
form:
\beq \label{GLRINT}
\frac{d N({\mathbf{x_{01}}},b_t,y)}{d y}\,\,\,=\,\,\,- \,\frac{2
\,C_F\,\as}{\pi} \,\ln\left( \frac{{\mathbf{x^2_{01}}}}{\rho^2}\right)\,\,
N({\mathbf{x}},b_t,y)\,\,\,+
\,\,\,\frac{C_F\,\as}{\pi}\,\,
\int_{\rho} \,\,d^2 {\mathbf{x_{2}}}\,
\frac{{\mathbf{x^2_{01}}}}{{\mathbf{x^2_{02}}}\,
{\mathbf{x^2_{12}}}}\,\,\times\,\,
\eeq
$$
\left(\,\,2\,N({\mathbf{x_{02}}},{ \mathbf{ b_t -
\frac{1}{2}
x_{12}}},y)\,\,\,-\,\,\,N({\mathbf{x_{02}}},{ \mathbf{ b_t - \frac{1}{2} 
x_{12}}},y)\,\,N({\mathbf{x_{12}}},{ \mathbf{ b_t - \frac{1}{2} 
x_{02}}},y)\,\,\right)\,\,.
$$
Assuming that ${\mathbf{x_{12}}}$ and ${\mathbf{x_{20}}}$ are much smaller
than the values of the typical impact parameter ( $b_t$ ), one can see
that \eq{GLRINT} is the GLR equation but in the space representation. It
turns out that this representation is very convenient for searching
solutions to  \eq{GLRINT}.

\begin{figure}[htbp]
\begin{center}
  \epsfig{file=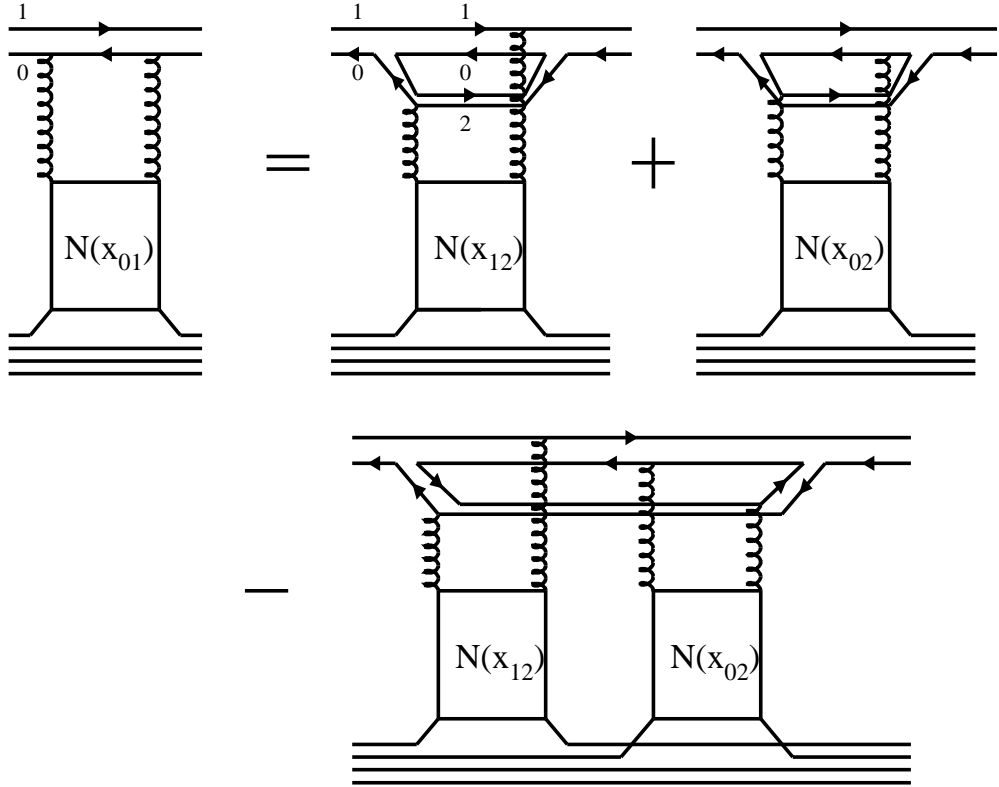,width=170mm}
  \caption[]{\it Pictorial representation of the nonlinear evolution
equation. }
 \end{center}
\label{fig2}
\end{figure}

The first term in the  r.h.s. of the equation gives the contribution
of
virtual corrections, which appear in the equation as a  result of the
normalization of the partonic wave function of the fast colour dipole (
see Ref. \cite{MUDIPOLE} ). The second term describes the decay of the
colour dipole with the size ${\mathbf{x_{01}}}$ into two dipoles wth sizes
${\mathbf{x_{02}}}$ and ${\mathbf{x_{12}}}$ and their  interactions 
 with the target in the impulse approximation ( notice factor 2 in
\eq{GLRINT} ). The third term corresponds to simultaneous interaction of
two produced colour dipoles with the target and describes the Glauber-type
corrections for scattering of these dipoles.

Being differential equation, \eq{GLRINT} needs an  initial condition at $y
=0$ to be solved. In Ref. \cite{KOV}it  was shown that the initial
condition
for  $N({\mathbf{x_{01}}},b_t,y )$ is the Mueller - Glauber formula
\cite{MU90}, namely,
\beq \label{MG}
N({\mathbf{x_{01}}},b_t,y=0
)\,\,\,=\,\,\,2\,\,\left(\,\,1\,\,\,-\,\,\,\exp\{\,
-\,\,\frac{\as
\,\pi^2}{4\,N_c\,\pi\,R^2_A}\,{\mathbf{x^2_{01}}}\,A\,\,
x_0G^{DGLAP}_N(x_0,\frac{1}{{\mathbf{x^2_{01}}}})\,\,
e^{-\frac{b^2_t}{R^2_A}}\}\,\,\right)\,\,;
\eeq
where $x_0G^{DGLAP}_N(x_0,\frac{1}{{\mathbf{x^2_{01}}}})$ is the gluon
structure function obtained as a solution to the DGLAP linear evolution
equations \cite{DGLAP}.

\subsection{ The strategy  of searching for solutions}

\eq{GLRINT} is nonlinear integro-differential equation which is rather
difficult to
solve. However, this equation can be simplified and can be  reduced to
nonlinear  but defferential equation in partial derivatives in two
different
kinematic regions (see Fig.1 ): to the right of the critical line ($r
\,>\,r_{cr} $) and to the left of the critical line ($ r\,<\,r_{cr}$).

\subsubsection{ $\mathbf{r\,\,>\,\,r_{cr}}$}
Indeed, for $ r\,>\,r_{cr} $, the size of the two produced colour dipoles
( $ {\mathbf{x_{02}}}$ and $ {\mathbf{x_{12}}}$ )
are much larger than the size of the initial colour dipole ($
{\mathbf{x_{01}}}$ ). Therefore, we can reduce the kernel of \eq{GLRINT}
to \cite{KOV}
\beq \label{DLR}
\int_{\rho} \,\,d^2
{\mathbf{x_2}}\,\,\frac{{\mathbf{x^2_{01}}}}{{\mathbf{x^2_{02}}}\,
{\mathbf{x^2_{12}}}}\,\,\,\longrightarrow\,\,\,{\mathbf{x^2_{01}}}
\,\pi\,\,\int^{
\frac{1}{\Lambda^2_{QCD}}}_{{\mathbf{x^2_{01}}}}\,\,\frac{d 
{\mathbf{x^2_{02}}}}{( \,{\mathbf{x^2_{02}}}\,)^2}\,\,.
\eeq

Introducing a new function $n({\mathbf{x_{01}}},b_t,y
)\,\,=\,\,N({\mathbf{x_{01}}},b_t,y )/{\mathbf{x^2_{01}}}$ and using the
fact that the virtual corrections does not contribute in the double log
approximation ( see Ref. \cite{KOV} for example ), one can obtain the
equation:
\beq \label{GLRDIFR}
\frac{d^2 n_{\rightarrow}(r,b_t,y)}{d y \,\,d
\,r}\,\,=\,\,\frac{\as\,C_F}{\pi}\,\,\{\,2\,n_{\rightarrow}(r,b_t,y)
\,\,\,-\,\,e^{- r}\,n^2_{\rightarrow}(r,b_t,y)\,\}\,\,;
\eeq
where $r\,=\,\ln(1/{\mathbf{x^2_{01}}})$.

In section 3 we will discuss  the solution to this equation which
satisfies
the initial condition of \eq{MG}.

\subsubsection{ $\mathbf{r\,\,< \,\,r_{cr}}$}
As have been discussed many times ( see for example Ref. \cite{LERY} and
Ref. \cite{MU99} ), the new scale appears in hdQCD which has a simple
physical meaning of the average transverse momentum of parton in the
parton cascade ( $Q_{cr}(x,b_t )$ ). In this region the size of the
initial colour dipole is larger than the typical size which we expect in
hdQCD parton cascade (  ${\mathbf{x^2_{01}}}\, >$ $1/Q_{cr}(x,b_t )$ ). 
Therefore, the  main contribution in \eq{GLRINT} comes from the
configuration  when one of the produced colour dipole has a size much
smaller than the size of the initial colour dipole \footnote{This idea
has been suggested in Ref.\cite{KOV} in one of the first version of this
paper but disappeared in the final version.}. We anticipate
that the size of the smallest colour dipole will be of the order of
$1/Q_{cr}(x,b_t )$. Such a configuration simplify the kernel in
\eq{GLRINT} which has a form
\beq \label{DLL}
\int_{1/Q_{cr}(x,b_t )} \,\,d^2
{\mathbf{x_2}}\,\,\frac{{\mathbf{x^2_{01}}}}{{\mathbf{x^2_{02}}}\,  
{\mathbf{x^2_{12}}}}\,\,\,\longrightarrow\,\,\,
\,\pi\,\,\int^{{\mathbf{x^2_{01}}}}_{1/Q_{cr}(x,b_t )}\,\,\frac{d
{\mathbf{x^2_{02}}}}{{\mathbf{x^2_{02}}}\,}\,\,+ \,\,\pi\,\,
\int^{{\mathbf{x^2_{01}}}}_{1/Q_{cr}(x,b_t )}\,\,\frac{d
{\mathbf{x^2_{12}}}}{{\mathbf{x^2_{12}}\,}}\,\,;
\eeq
The sum of two terms reflects the fact that two different
produced  colour dipoles can be small.
 In momentum representation \eq{DLL} means that we sum $\log^n
(Q^2_{cr}(x,b_t)/Q^2)$ which are the normal contributions for the DGLAP
evolution coming from integration over transverse momentum from the small
transverse momentum ($Q^2$ ) to the large one ( $
Q^2_{cr}(x,b_t )$ ).  

  Introducing a new function
$\tilde{N}(r,b_t,y)\,\,=\,\,\int_r\,d r'
N(r',b_t,y)$, we reduce \eq{GLRINT} to the differential equation:
\beq \label{GLRDIFL}
-\, \frac{d^2
 \tilde{N}_{\leftarrow}(r,b_t,y)}{d\,y\,\,d\,r}\,\,\,=\,\,\,
\frac{2\,\as\,C_F}{\pi}\,\,\{\,\tilde{N}_{\leftarrow}(r,b_t,y)\,\,+ \,\,
{\tilde{N}}_{\leftarrow}(r,b_t,y)\,\frac{ d
{\tilde{N}}_{\leftarrow}(r,b_t,y)}{d r}\,\}\,\,.
\eeq
One can see that in differential equation ( \eq{GLRDIFL} ) we lost any
dependence of the scale $Q_{cr}(x,b_t )$ and it will come back to the
problem only in matching a solution to \eq{GLRDIFL} with a solution to
\eq{GLRDIFR}.

In section 4 we will find a solution to this equation which match the
solution of \eq{GLRDIFR} on the critical line.

\subsubsection{The general solution}

Let us summarize what solution we are looking for:
\begin{enumerate}
\item\,\,\,We are going to solve the differential equation (see
\eq{GLRDIFR}
) to the right of the critical line which satisfies the initial
condition given by the Mueller-Glauber
formula (see \eq{MG};

\item\,\,\, We are going to find the solution to the differential equation
( see \eq{GLRDIFL} ) to the left  of the critical line;

\item\,\,\,We match the solution of \eq{GLRDIFR} with the solution of
\eq{GLRDIFL} on the critical line;

\item\,\,\,We
can hope that the solution of these two differential equations  will be
close to the solution to \eq{GLRINT} in the vicinity of the critical line
($ r\,\approx\,r_{cr} $)since we provide the matching of these two
solutions on the critical line.
\end{enumerate}

\subsection{Experience of solving the GLR-type nonlinear equations}
During the past two decades we have learned the main properties of the
solution to
the GLR-like nonlinear differential equations. Our experience in solution
of these equations based on three different approaches that have been
developed. 

First, these equations were solved in semiclassical approximation,
assuming $N = e^S$ where $S$ is a smooth function of $r$ and $y$, namely
$d^2 S/d y d r \,\,\ll\,\,( dS/dy )^2 $ or $ ( d S/d r )^2$ (see Refs.
\cite{GLR}\cite{BART} \cite{COKW}\cite{AGL} and references therein). In
this approximation the GLR - type equation can be solved using
characteristics method which leads to existence of a  special ( critical
) line. All charecteristics in the region of low $x$ cannot cross this
line but can approach it.  The nonlinear term provides that
  the parton density $N(r,b_t = 0,y)$    is constant on critical line.   
To the right of the critical line ( $r \,>\,r_{cr}$ ) the nonlinear
corrections turns out to be small and the solution can be found as a
solution of the linear DGLAP evolution equation but with the boundary
conditions: $N = Const $ on the critical line.  For our attempts to find a
solution the important message from the semiclassical approach is the fact
that we have a new scale ($Q_{cr} (x,b_t )$ ) and the properties of the
solution looks differently for $r > r_{cr} = \ln Q^2(x,b_t)$ and for $r <
r_{cr}$.

In Ref. \cite{LALE} a new method was suggested for solving \eq{GLRDIFR}.
The generating function was constructed which linearize
 \eq{GLRDIFR} reducing \eq{GLRDIFR} to a linear equation in partial
derivatives but with one more variable. It is easy to find a general
solution of this linear equation and all difficulties are concentrated  
in finding  the solution which satisfies the boundary condition.
We are going to use this method to find a solution to the right of the
critical line. 

In the region of very low $x$,  a new idea for searching  a solution of
the
GLR-type nonlinear equation was suggested in Ref.\cite{BALE}. In this
paper it was argued that the solution of \eq{GLRDIFR}, which satisfies all
 physics restrictions , is a function of one
variable: $z = 4
\bar \as y \,- \,r $. We will find this solution in an explicit way in
section 4 and will show how to match this solution with the solution to
the right of the critical line.

\subsection{Theory status of the equations}

In this subsection we recall the main assumptions that have been made
to obtain the nonlinear evolution equationsfor hdQCD (see \eq{GLRINT},
\eq{GLRDIFR} and \eq{GLRDIFL} ).
\begin{enumerate}
\item\,\,\,For derivation of all equations we used the leading log(1/$x$ )
approximation of pQCD. Only in this approximation we can neglect the
energy recoiled during emission of the colour dipole ( the linear term in
equations ) or during rescattring of this dipole ( the nonlinear term ).
Formally speaking, we have to assume that parameter $\as
\log(1/x)\,\,\geq\,\,1$ while $\as\,\,\ll\,\,1 $; 
\item\,\,\,To obtain the differential equations ( see \eq{GLRDIFR} and
\eq{GLRDIFL} ) we have to  assume  existence of a more restricting
parameter $\as r y
\,\geq\,\,1$, but $\as r \leq 1$ and $\as y \leq 1$,   which corresponds
to the double log approximation of pQCD in
the linear evolution equation;

\item\,\,\, The number of colours should be large ( $N_c \,\gg\,1 $). It
has been shown\cite{DYNCOR}  that for $N_c \,\approx\,1$ the  correlations
between
colour
dipoles should be taken into account which break the simple physical
picture of Fig.2.  In the double log approximation these correlations can
be taken into account \cite{LALE} but they lead to rather complicated
equations. On the other hand, in Ref. \cite{LALE} it was demonstrated that
all these complications lead to corrections of the order of $1/N^2_c $ 
and can be neglected even for $N_c$ = 3;

\item\,\,\,The master equation ( see \eq{GLRINT} ) sums so called ``fan"
diagrams ( see Fig. 3a ), which reflect the fact that the fast colour
dipole decays into two colour dipoles.  However, the recombination of two
colour dipoles into one have been omitted (see Fig. 3b). For $r > r_{cr}$
it has been proven that such diagrams give only small corrections
\cite{GLR}  but in
the kinematic region to the left of the critical line  such enhanced 
diagrams  can be neglected only in the case of scattering off the 
heavy  nucleus. It turns out ( see Refs. \cite{SCHWIM} \cite{AGL}
\cite{KOV} )
that in the case of heavy nucleus the contributions of ``fan" diagrams are
proportionsal to $\gamma A^{\frac{1}{3}}$ while enhanced diagrams of
Fig.3b are of the order of $\gamma$ without large factor
$A^{\frac{1}{3}}$. Therefore, our approach is valid only for heavy nuclei
and can be considered only as a model for the interaction with a hadron;

\item\,\,\,We assume that (i) there are no correlations between different
nucleons in a nucleus and (ii) the average $b_t$ for colour dipole -
nucleon interaction is much smaller than $R_A$. Both these assumptions
are usual for treatment of nucleus scattering.
\end{enumerate}

\begin{figure}[htbp]
\begin{tabular}{ c c}
\epsfig{file=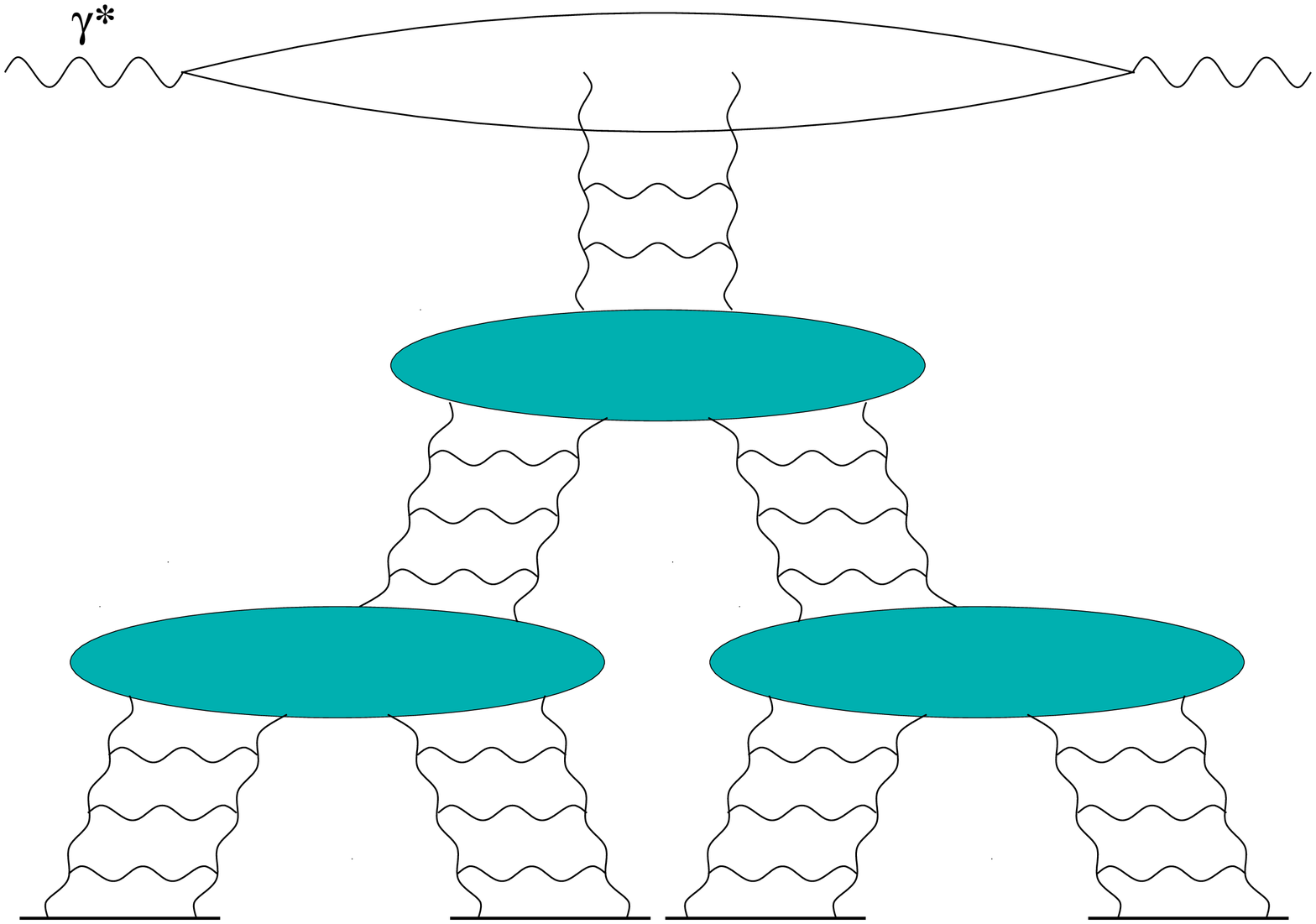,width=90mm,height=60mm} &
\epsfig{file=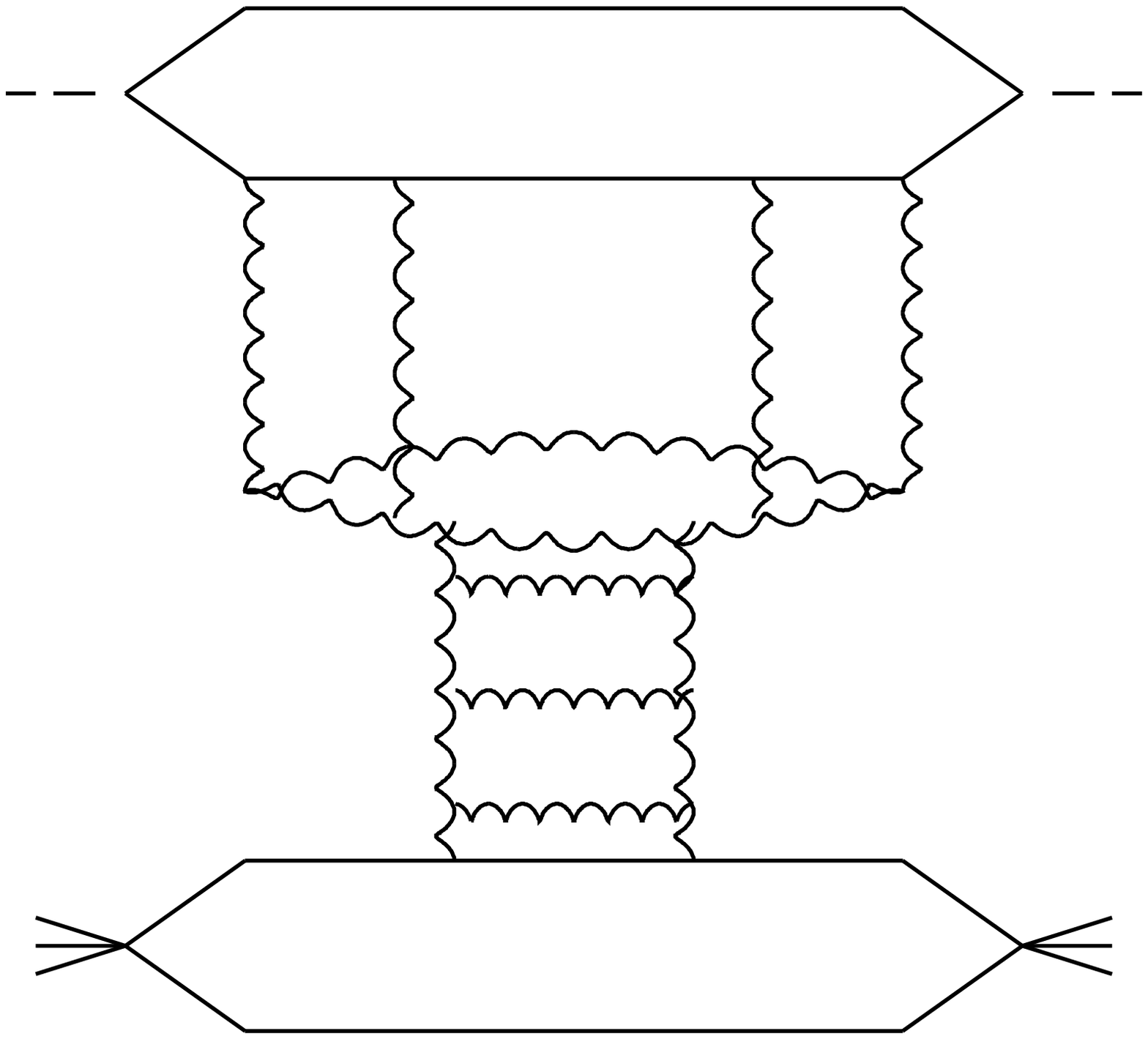,width=80mm,height=60mm}\\
Fig. 3-a & Fig.3-b \\
\end{tabular}
  \caption{\it ``Fan" ( Fig. 3a ) and enhanced ( Fig. 3b ) diagrams. }
\label{fig3}
\end{figure}

\section{Solution to the right of the critical line }
\setcounter{equation}{0}
In this kinematic region we   use the generating function method, proposed
in Ref. \cite{LALE}, which linearize \eq{GLRDIFR}.  Following Ref.
\cite{LALE}, we introduce a  generating function
\beq \label{GENFUN}
g_A(r,y,b_t,\eta)\,\,\,=\,\,\,\sum^{\infty}_{n =
1}\,\,\,\,n^{(n)}_{N,\rightarrow}(r,y,b_t)\,e^{n\,\eta}\,\,,
\eeq
where $n^{(1)}_{\rightarrow}\,=\,n_{\rightarrow}(r,y,b_t) =
N(r,b_t,y)\,e^r \,=\,(\as \,\pi^2/2 N_c)\,\,\rho(r,b_t,y)$ is proportional
to the density of colour dipoles in the target. $n^{(n)}_{\rightarrow}$ is
proportional to the probability to have $n$ - colour dipoles with the same  
size 
$e^{-  r }$ in the parton cascade. In the case, when we neglect 
correlations between colour dipoles in the parton cascade, 
$n^{(n)}_{\rightarrow}\,\,=\,\,C_n(A) 
\left(\,n^{(1)}_{\rightarrow}\,\right)^n$.
Coefficients $C_n(A)$ should be found from the initial condition of
\eq{MG}.

For $n^{(n)}_{\rightarrow}( r,b_t,y)$ we can write the evolution equation
of the \eq{GLRDIFR} - type, noticing that  every colour dipole with size
$r$ at rapidity $y$ and impact parameter $b_t$ can either propagate to
rapidity $y + d y $ changing its size from $r$ to $r + d r$ ( see first
two terms in Fig.2 )  or decay
into two colour dipoles ( see the last term in Fig. 2  and Fig. 3a). This
observation leads to
equation \cite{LALE}
\beq \label{EQGF}
\frac{\partial^2 n^{(n)}_{\rightarrow}(r,b_t,y)}{\partial y \partial
r}\,\,\,=\,\,\,\bar \as \,\{\,n^2\,\,
n^{(n)}_{\rightarrow}(r,b_t,y)\,\,\,-\,\,\,n\,\,e^{-r}\,\,
n^{(n+1)}_{\rightarrow}(r,b_t,y)\,\}\,\,,
\eeq
  
where coefficient $\bar \as \,n^2$  corresponds to $\omega
\,\gamma_n(\omega)$
where $\gamma_n$ is the anomalous dimension of $n^{(n)}_{\rightarrow}$ and
$\omega $ is variable conjugated to $y$. In Refs.\cite{ANDIM} the limit of
small $\omega$ ( low $x$ )   for $\gamma_n $ has been calculated. 
 
Comparing \eq{GENFUN} with \eq{MG} one can obtain the following
information on the generating function of \eq{GENFUN}:
\begin{eqnarray}
&
C_n \,\,\,=\,\,\,\frac{( -1 )^{n + 1}}{2^n\,\,n!}\,\,; &\label{INCONGF1}\\
&
N(r,b_t,y)\,\,\,=\,\,\,2\,
\,g_A\left(\,r,y,b_t,\eta\,\,=\,\,-r
\,+\,\eta_A\,\right) \,\,;&\label{INCONGF2} \\
&
n^{(n)}_{N,\rightarrow}(r,y 
= 0 (x = x_0),b_t)\,\,=\,\,C_n\left(\,n^{DGLAP}_{N,\rightarrow}(r, y = 0 
( x = x_0 ),b_t \,\right)^n\,\,;&\label{INCONGF3}\\
&
n^{DGLAP}_{N,\rightarrow}(r, y = 0 (x = x_0 ),b_t)\,\,\,=\,\,\frac{\as
\,\pi^2}{2 N_c \pi R^2_N}\,\,\,\,x_0
G_N(x_0,e^r)\,S_N(b_t)\,\,;&\label{INCONGF4}\\
& 
\eta_A\,\,\,\,=\,\,\,\ln\left(\,A\,R^2_N\,S_A(b_t)\,\right)\,\,\,.&
\label{ETAA}
 \end{eqnarray}
\eq{INCONGF3} and \eq{INCONGF4} mean that we consider $x = x_0$ to be so
small that we can neglect correlations between produced colour dipoles.

\eq{EQGF} can be rewritten as a linear equation for the generating
function
$g_A$ of \eq{GENFUN}:
\beq \label{EVGF}
\frac{\di^2}{\di y\di r} g_A (r,y,b_t,
\eta,)\,\,\,=
\eeq
$$
\baralpha\frac{\di^2}{\di\eta^2}
g_A( r,y,b_t, \eta )\,\,-\,\,\baralpha \gamma e^{ - r -\eta}\,\,
\left(\,\frac{\di g_A( r,y,b_t, \eta )}{\di\eta}-g_A( r,y,b_t, \eta ) 
\,\right)\,\,,
$$
with $\gamma = 1 $.

This equation is a linear equation which can be solved just going to
Mellin transform with respect to $y$ and $\xi = r + \eta$:
\beq \label{MELLIN}
g_A(\xi, y,b_t,\eta )=\int\frac{d\omega dp}{(2\pi i)^2}
g_A(\xi,\omega,b_t, p)\,e^{\omega y+p\eta}.
\eeq
In \eq{MELLIN} the contours of integration over $\omega$ and $p$ 
 lie  along the imaginary axis to the right of all singularities in
$\omega$ and $p$.
 
Substituting \eq{MELLIN} into \eq{EVGF} we obtain the following equation
for  the Mellin image
  $g_A(\xi,\omega,b_t, p)$.
\beq \label{EVGFMEL}
\frac{d g_A(\xi,\omega,b_t, p)}{d \xi
}\,\,\,=\,\,\frac{\baralpha}{\omega}\,\,\,\{\,\,
p^2\,\,-\,\,\gamma\,(\,p - 1 \,)\,\,e^{- 
\xi}\,\,\}\,\,g_A(\xi,\omega,b_t, p)\,\,,
\eeq
which has an obvious solution:
\beq \label{EVGFMELSOL} 
g_A(\xi,\omega,b_t,p)\,\,\,=
\,\,g_A(\omega,p,b_t)\,\,e^{\frac{\baralpha}{\omega}\,p^2\,\,\xi
\,\,+\,\,\frac{\baralpha \,\gamma}{\omega}\,(  p - 1 )\,\left(\,e^{- \xi}
\,-\,e^{- \eta_A}\,\right)}\,\,.
\eeq
Function $g_A(\omega,p,b_t)$ should be found from the initial condition of
\eq{MG} ( see also \eq{INCONGF1} - \eq{INCONGF4} ).  
Our statement is that 
\beq \label{MELINCON}
g_A(\omega,b_t,p)\,\,\,= \,\,\,\Gamma(-f(p))\,\,\,
\left( \frac{\as \pi^2}{4\pi R^2_N} \right)^f\,\,
\frac{df(p)}{dp}e^{\frac{\baralpha
\gamma}{\omega}\,(p - 1)\,e^{ -
\eta_A}}\,\,e^{\frac{\baralpha}{\omega}\,p^2\,\,\eta_A} \eeq
satisfies initial conditions, with $f(p) = p
\,\,+\,\,\frac{\baralpha\,p^2}{\omega}$. Substituting \eq{MELINCON} in
\eq{EVGFMELSOL} and changing variables of integration from $p$ to $f$ ,we 
obtain
\beq \label{EVGFSOL}
g_A( \xi,b_t,y,\eta )=\int \frac{d\omega df}{( 2\,\pi\,i )^2}\,\,
\Gamma(- f)\,\,\left( \frac{\as \pi^2}{\pi R^2_N} \right)^f
e^{- p(f) r  \,+\,f\xi \,+ \,f\eta_A +  \omega y+\frac{\baralpha \gamma
\,( p(f) - 1
)}{\omega}}\,\,,
\eeq
where $p(f)$ is determined by
\beq \label{F}
f\,\,=\,\,p ( f )\,\,+\,\,\frac{\baralpha}{\omega}\,p^2(f)\,\,.
\eeq

Indeed, at $y = y_0  $ ( $x = x_0$ ) , $| \xi | \,\,\ll\,\,r$, $\xi=
\eta_A$, $r \,\,\gg\,\,1$, we have

\begin{eqnarray} 
g_A( r,b_t,y= y_0, \eta = - r + \eta_A ) &=& 
\int \frac{d\omega df}{(2\,\pi\,i)^2}\,
\left( \frac{\as \pi^2}{4\,N_c \pi R^2_N} \right)^f\,\,\Gamma( - f )\,
e^{- f\,r \,+\,f \,\eta_A\,\,+\,\,\omega\,
y_0\,\,-\,\,\frac{\baralpha}{\omega}\,f^2}\nonumber\\
  &=& \sum_{n=1}^{\infty}\frac{(-1)^n}{n!}\,e^{ -n \,r}\,\left( \frac{\as
\pi^2}{N_c\pi R^2_N} \right)^n
  e^{n\eta_A}\,\int\frac{d\,\omega }{2\pi i} \, .\,e^{\omega
  y_0-\frac{\baralpha n^2 r}{\omega}}\,\,\nonumber\\
  &=& \sum_{n=1}^{\infty}\frac{(-1)^n}{n!}\,\left(\,\frac{\as
\pi^2}{\,4\,N_c\,\pi \,R^2_A}\,{\mathbf{x^2_{01}}}\,A x_0 G^{DGLAP}_N(
x_0,\frac{1}{{\mathbf{x^2_{01}}}})\,\right)^n \,\,.\label{CHECKIN}
\end{eqnarray}

Taking into account \eq{INCONGF2}, one can see that \eq{CHECKIN}
reproduces correctly the initial condition of \eq{MG}.

It should be stressed that the contour of integration over $f$ was taken
along positive real axis in  such a way that all positive and integer $f$
are enveloped by it.  The integral over $f$  is well defined due to rapid
decrease
of $\Gamma( - f )$ ( see Ref. \cite{LALE} ). To study asymptotic behaviour
of the parton densities at low $x$ we have to move the contour toward the
imaginary axis. The asymptotic stems from the rightmost singularities in
$f$ as well as from possible saddle points in this integral. The first
dangerous region is a vicinity of $f \,\rightarrow\,1 $ where we expect a
saddle point. To investigate a situation near f$ = 1 $, we introduce a new
variable  $f \,=\,1\,\,+\,\,t$, hoping that $t$ will be very small.
At $t\,\rightarrow \,0$ our solution of \eq{EVGFSOL} has a form
\begin{eqnarray}
g_A (r,b,y,\eta = \eta_A + \tilde{\eta}) &=&
\int\frac{d\omega df}
{ (2\pi i)^2 }\Gamma(-f)e^{\frac{\baralpha \gamma}
{\omega}(p-1)\,e^{-\eta_A } +f (\eta_A + \tilde{\eta}) + pr+\omega
y}\nonumber\\
&=& \int \frac{d\omega\omega dt}
{(2\pi i)^2 t}e^{\psi(\omega,r,t,b_t)}
\,e^{\eta_A + \tilde{\eta}}\,\,,\label{SADDLE}
\end{eqnarray}

where $\tilde{\eta}\,=\,\ln\left( \frac{\as \pi^2}{N_c \pi R^2_N}
\right)$.

In Appendix A we calculate the integral of \eq{SADDLE} using the saddle
point approximation and the answer, that we obtain in this appendix (
see \eq{SOLCRLINE} )  can be written in the form:
\beq \label{SCRL}
g_{A, \rightarrow}(y,\xi=\eta_A,r,b_t)\,=\,\frac{\omega^2_{cr}}{2 \pi\,
2\baralpha (r) \,t_0\,\varsigma}\,\frac{ \as \pi^2 A}{4 \pi \,R^2_A}
\,e^{-
\frac{b^2_t}{R^2_A}}\,
e^{\omega_{cr}y - r+ \frac{\baralpha r }{\omega_{cr}}} \,\,,
\eeq
with $\omega_{cr}$ is given by \eq{OMEGACR} and $t_0$ by \eq{OMSAD}.

One can see that \eq{SCRL} leads to $g_A(y,\xi=\eta_A,r,b_t)$ which is
constant on the critical line given by equation:
\beq \label{EQCRLINE}
\tilde{\Psi}\,\,=\,\,\omega_{cr}\,y_{cr}\,\,-\,\,r
\,\,+\,\,\frac{\baralpha}{\omega_{cr}}\,r
\,\,+\,\,\eta_A\, + \,\tilde{\eta}\,\,-\,\,\ln\frac{2 \baralpha
\,\varsigma\,r\,t_0}{\omega_{cr}}\,\,\,=\,\,Const(\as)
\eeq
where $Const(\as)$ can be a function of   $\as$ but it
does not depend on neither $y$ nor $r$ or $\eta_A$. We anticipate that the
value of
$t_0$ will be small on the critical line and, because of this, 
the last term in \eq{EQCRLINE} has been taken into account in the phase
$\tilde{\Psi}$. 

The solution of \eq{EQCRLINE} gives
\beq \label{CRLINEFIN}
y_{cr}\,\,=\,\,\frac{1}{4 \baralpha}\,\left(\,r\,\,-\,\,2( \eta_A\, +
\,\tilde{\eta} )\,\, +\,\,2\,\right)\,\,.
\eeq
This line gives the minimal value for $\tilde{\Psi}$, namely,
$\tilde{\Psi}\,=\,1$.

The value of $
g_A(y,\xi=\eta_A,r,b_t)$ is equal to
\beq \label{VALCR}
g_{A, cr}(y,\xi=\eta_A,r,b_t)\,\,\,=\,\,\,\frac{e\,\,\baralpha}{\pi}.
\eeq

Solving \eq{CRLINEFIN} with respect to $r$ we obtain the typical momentum 
( colour dipole size ) on the critical line
\beq \label{Q0X}
Q^2_{cr}(x;b_t)\,\,=\,\,Q_0^2 \left(\frac{\as}{N_c
 R^2_N\mathbf{e}}\right)^2
\,\,
\left( \frac{A R^2_N}{R^2_A}\right)^2 \,\,
e^{4 \baralpha y}\,\,e^{- \frac{2 b^2_t}{R^2_A}}.
\eeq

At $b_t = 0 $ $Q^2_{cr}(x;b_t=0) \,\,\propto\,\,A^{\frac{2}{3}}$, as can
be see easily from \eq{Q0X}.

Comparing \eq{VALCR} with \eq{NVSG} one sees that
$xG_A(x,\frac{1}{{\mathbf{x^2}}})$ on the critical line is equal to 
  
\beq \label{XGCRR}
xG_{A, cr}
(x,\frac{1}{{\mathbf{x^2}}})\,\,\,=\,\,\,\frac{2 N_c \pi R^2_A
Q^2_{cr}}{\as \pi^2 }\,N({\mathbf{x^2}},b_t=0,y_{cr})\,\,=\,\,
{\frac{4 N_c^2 e}{\pi^3}}
\,R^2_A\,Q^2_{cr}(x)\,\,\propto\,\,A^{\frac{4}{3}}\,\,. 
\eeq
 It is interesting to notice that such A-dependence can be understood
directly from the Mueller - Glauber formula of \eq{MG}. Indeed, the
shadowing corrections become to be essential when 
\beq \label{MGEST}
\frac{\as \pi^2}{4 N_c \pi R^2_A}{\mathbf{x^2}}  \,A x
G(x,1/{\mathbf{x^2}})\,\,\approx\,\,1\,\,.
\eeq
Recalling that anomalous dimension  in the vicinity of the critical
line is equal to $\gamma(\omega) =
\frac{\baralpha}{\omega_{cr}}\,=\,\frac{1}{2}$, one can find from
\eq{MGEST} that
${\mathbf{x^2_{cr}}}\,\,=\,\,1/Q^2_{cr}\,\,\propto\,\,A^{-\frac{2}{3}}$,
which is the same A-dependence that we obtained from more advanced
calculations ( see \eq{XGCRR} ).  

 \section{Solution to the left of the  critical line}
\setcounter{equation}{0}
In this section we are going to solve \eq{GLRDIFL} using the idea
suggested in Ref. \cite{BALE}, namely, $\tilde{N}_{\leftarrow}\,\,=\,\,
\tilde{N}_{\leftarrow} (z) $ where 
\beq \label{Z}
z\,\,\,=\,\,\,4\,\baralpha
 \,\,y\,\,\,-\,\,\, r\,\,- \,\,\beta(A;b_t)\,\,.
\eeq
Function $\beta(A;b_t)$ can be chosen in a such way that the equation for
the critical line in terms of new variable $z$ looks as
$z = 0 $.  One can see from \eq{CRLINEFIN} that $\beta(b_t) $ is equal to 
\beq \label{BETA}
\beta(b_t)\,\,\,=\,\,\,- 2\tilde{\eta}
\,\,-\,\,2\,\ln\frac{A\,R^2_N}{\pi R^2_A} \,\,+\,\, \frac{2 b^2_t}{R^2_A}
\,\,+ \,\, 2\,\,. 
\eeq

Assuming that  $\tilde{N}_{\leftarrow} (r,b_t,y)$ is a function of $z$
and $b_t$ only, we can rewrite \eq{GLRDIFL}
in the form:
\beq \label{EQZL}
\frac{d^2 \tilde{N}_{\leftarrow}(z,b_t)}{dz
^2}\,\,\,=\,\,\frac{1}{4}\,\,\left(\,1\,\,\,-\,\,\,\frac{d
\tilde{N}_{\leftarrow}(z,b_t)}{d z
}\,\right)\,\,\tilde{N}_{\leftarrow}(z,b_t)
\,\,.
\eeq
Here we used the explicit equation for variable $z$, namely, $ z =
4\baralpha  y - r  - \beta(b_t)  $. 
\subsection{A solution to \eq{EQZL}}

First, we find a solution to \eq{EQZL} introducing function $\zeta(z)$ in
the following way:
\beq \label{ZETA}
\tilde{N}_{\leftarrow}(z,b_t) \,\,=\,\,\int^z\,\,d z'
\,\left(\,1\,\,-\,\,e^{
- \zeta(z,b_t)}\,\right)\,\,.
\eeq
For function $\zeta(z)$ \eq{EQZL} can be reduced to 
\beq \label{EQZ}
\frac{ d \zeta(z,b_t)}{d z} \,\,\,=\,\,\,\frac{1}{4}\,\,\int^z\,\,d z'
\,\,\left(\,1\,\,-\,\,e^{- \zeta(z',b_t)}\,\right)\,\,.
\eeq
Changing variable in the integral from $z'$ to $\zeta'$ we have
\beq \label{ZETA1}
\frac{ d \zeta(z,b_t)}{d z}\,\,\,=\,\,\,\frac{1}{4}\,\,\int^{\zeta}_
{\zeta_0(b_t)}\,\,\frac{d
\zeta'}{\frac{d \zeta'}{d z'}} \,\left(\,1\,\,-\,\,e^{-
\zeta'}\,\right)\,\,.
\eeq
\eq{ZETA1} can be easily solved and the solution is
\beq \label{ZETASOL}
z\,\,=\,\,\sqrt{2}\,\int^{\zeta}_{\zeta_0(b_t)} \,\,\frac{ d
\zeta'}{\sqrt{\zeta'\,\, +\,\, \,(\,e^{- \zeta'}\,\,-\,\,1\,)}}
\,\,.
\eeq
The numerical solution to \eq{ZETASOL} is given in Fig.4. 
This solution depends on the initial conditions on the critical line .
There are two of them,
namely,
\begin{eqnarray}
&
N_{\leftarrow}(z = 0 ) = N_{\rightarrow}(z=0) \,\,;&\label{INCON1}\\ 
&\frac{d \ln N_{\leftarrow}}{d z}|_{z =0}\,\,=\,\,\frac{d \ln
N_{\rightarrow}}{d z}|_{z =0}\,\,=\,\,\frac{1}{2}\,\,.& \label{INCON2}
\end{eqnarray}

We can assume that $\zeta \,\,\gg\,\,1$ at large $z$. In this
case we can neglect $e^{- \zeta'}$ in the integrand and we obtain an
explicit analytic solution:
\beq \label{ZETAANSOL}
\zeta(z) \,\,=\,\,(\,\sqrt{\zeta_0(b_t)}\,\,+\,\,\frac{z}{2\sqrt{2}}\,)^2\,\,,
\eeq
or in terms $N(z,b_t)$ we have
\beq \label{ZETAN}
N_{\leftarrow}(z,b_t)\,\,\,=\,\,\,1\,\,\,-\,\,\,\,
e^{-\,\,(\,\sqrt{\zeta_0(b_t)}\,\,+\,\,\frac{z}{2\sqrt{2}}\,)^2}\,\,\,
\eeq
where $z_0(b_t)$ should be found from matching of this solution  with the
solution of the previous section on the critical line.
\subsection{Matching with the solution to the right of the critical line}
We cannot use the simple solution of \eq{ZETAANSOL} for matching
with the solution of \eq{SCRL} since $\zeta$ is expected to be rather
small in the vicinity of the critical line and, therefore, we cannot
neglect $e^{- \zeta}$ in \eq{ZETASOL}.  However, for small
$\zeta\,\,\ll\,\,1$ we easily obtain a simple solution to \eq{ZETASOL},
namely,
\beq \label{CRLINESOL}
\ln\frac{\zeta}{\zeta_0(b_t)}\,\,=\,\,\frac{1}{2}\,z
\,\,.
\eeq
\eq{CRLINESOL} leads to \eq{INCON2} of matching condition. This equation
allows us to find $\zeta_0(b_t)$, namely,
\beq \label{MATCHF}
1 \,\,-\,\,e^{- \zeta_0(b_t)}\,\,=\,\,N_{cr,\rightarrow }\,\,.
\eeq 
\eq{CRLINESOL} and \eq{MATCHF} provide matching as well as determine the
parameters of the asymptotic behaviour for $N_{\leftarrow}$ given by
\eq{ZETAN}. Using \eq{VALCR} one can find from \eq{MATCHF} that
\beq \label{VALZ}
\zeta_0(b_t)\,\,=\,\,2\,\frac{e\,\baralpha}{\pi}\,\,.
\eeq

\begin{figure}[htbp]
\begin{center}
  \epsfig{file=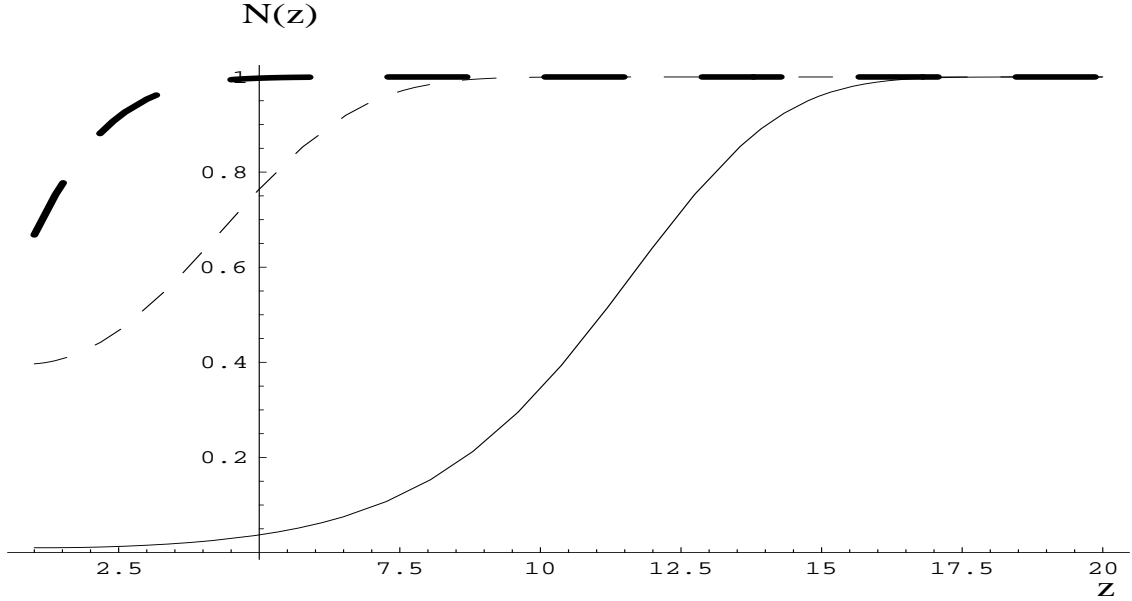,width=160mm, height=80mm}
  \caption[]{\it Dipole number density $N(z)$ as a function of
the critical  line parameter $z$. Solid line is the  numerical solution
with $\tilde{N}'(0)=0.01$, while   dashed line corresponds to
$\tilde{N}'(0)=0.4$. The bold dashed line shows the asymptotic solution
of \eq{ZETAN} with $\tilde{N}'(0)=0.4$.} 
 \end{center} 
\label{fig4}
\end{figure}

Fig.4 illustrates the behaviour of $N(z)$ versus  $z$. One can
see that   asymptotic is reached only at very large $z
\approx\,7 - 10$ or for very low $x$ ( $ x \,\approx\,10^{-6} - 10^{-7}$
).
At HERA we have $z \,\approx\,4 -5$ and we are far away from the
asymptotic solution. However,  we can
penetrate the region of large $z$ using a nuclear target,as one can see in
\eq{CRLINEFIN}.  Indeed, for
heavy nuclei we can easily have $2 \eta_A \,\approx\,3.5 - 4 $ . It gives
a possibility to have even at HERA kinematic region sufficiently large
values of $z\,\approx\, 7.5 - 9$.

\subsection{Stability of the solution}
 Arguments of the last section show, that
the solution given by \eq{ZETASOL} has a big chance to be the
solution to our equation. To prove this statement we need only to show
that the solution of \eq{ZETASOL} is stable with respect to small
perturbations of the initial conditions on the critical line. 
In other words, let us consider a small variation of the initial
conditions $| \delta \zeta_{cr} (y,r,b_t) |\,<\,\delta $ where $
0<\,\delta\,\ll\,1$. The solution is stable if for every $\delta$
we can found a small $\epsilon$  such that $\epsilon\,\,\rightarrow\,\, 0
$ at
$ \delta\,\,\rightarrow\,\,0$. We can find a linear equation 
 for function $\delta \zeta (y,r,b_t)$ substituting
$\zeta(y,r,b_t)\,\,\longrightarrow\,\,\zeta_s\,\,\,+\,\,\,\delta
\zeta(y,r,b_t)$ where $\zeta_s$ is given by \eq{ZETASOL}. Assuming that
$\delta \zeta_{cr} (y,r,b_t)$ is small we obtain a linear equation
\beq \label{STABEQ}
\frac{\di^2 \delta \zeta (y,r,b_t)}{\di \,y\,\,\di
r}\,\,=\,\,\baralpha\,\,\delta \zeta (y,r,b_t)\,\,e^{- \zeta_s(z)}\,\,.
\eeq
 In variables $z$ and $r$ this equation has a form
\beq \label{STABEQ1}
\frac{\di^2 \delta \zeta (z,r,b_t)}{\di \,z\,\,\di
r}\,\,=\,\,\frac{1}{4}\,\,\delta \zeta (z,r,b_t)\,\,e^{-
\zeta_s(z)}\,\,,
\eeq
which can be solved going to Mellin transform with respect to variable $r$.
Indeed, for Mellin image we have
\beq \label{MELSTAB}
\frac{d \delta \zeta (z,\nu,b_t)}{d \,z}\,\,=\,\,\frac{1}{4 
\,\nu}\,\,\delta \zeta (z,\nu,b_t)\,\,e^{-
\zeta_s(z)}\,\,.
\eeq
The general solution of \eq{MELSTAB} looks as follows
\beq \label{GENSOLST}
\delta  \zeta (z,\nu,b_t)\,\,\,=\,\,\int^{a + i \infty}_{a - i
\infty} \,\,\frac{d \nu}{2 \,\pi\,i}\,\delta\zeta(\nu) e^{\nu r
\,\,+\,\,\frac{1}{\nu}\,\int^z_0\,d z' \,\,e^{- \zeta_s(z',b_t)}}\,\,.
\eeq
The integral over $z'$ in \eq{GENSOLST} does not depend on $z$ at large
$z$. It means that the solution given by this equation is actually the
function of $r$ only. It is obvious that such a solution cannot be
tolerated by the initial conditions on the critical line.

\subsection{Low $\mathbf{x}$ ( high energy ) asymptotic}

As we have discussed the total dipole cross section is equal to
\beq \label{DICR}
\sigma_{dipole}\,\, =
\eeq
$$ 
\int\,d^2 b_t \,2
N(r,y,b_t)\,\,\,=\,\,2 \pi \,\{\int^{b^2_0}_0 d
b^2_t\,\,N_{\leftarrow}\,\,+\,\,\int^{\infty}_{b^2_0} \,\,d
b^2_t\,\,N_{\rightarrow}\,\,\}=\,\,2\pi 
\int^{b^2_0}_0 \,d b^2_t \,\,=\,\,2 \pi\,b^2_0 (y,r)
$$
since $N_{\leftarrow}(r,y,b_t)\,\,\longrightarrow\,\,\,1$ at low $x$ (
see \eq{ZETAN} ) and $N_{\rightarrow} (t,y,b_t)$ gives a small
contribution to the integral in \eq{DICR}.  

The
value for $b^2_0(y,r)$ we can evaluate recalling that solution of
\eq{ZETAN} is written for $z = 4\,\baralpha \,y - r + 2\tilde{\eta} +
2\ln\frac{A R^2_n}{R^2_A}\,- \,\frac{2
b^2_t}{R^2_A}\,\,>\,\,0$. Therefore, 
\beq \label{B0}
 b^2_0(y,r)\,\,=\,\,\frac{R^2_A}{2}\,\,\left(\,4\,\baralpha \,y -
r \,+\,2\ln\frac{A R^2_N}{R^2_A}\,\right)\,\,,
\eeq 
which gives
\beq \label{DICR1}
\sigma_{dipole}\,\,\,=\,\,\,\pi\,R^2_A\,\left(\,4\,\baralpha \,y -
r\,+\,2\,\ln\frac{A R^2_N}{R^2_A}\right)\,\,.
\eeq

For the gluon structure function \eq{DICR1} leads to
\beq \label{GAN}
x G(x,Q^2)\,\,\,\longrightarrow\,\,Q^2\,R^2_A\,\,\left(\,4\,\baralpha
\,\ln(1/x)\,\,  -
\,\,\ln Q^2\,+\,2\,\ln\frac{A R^2_N}{R^2_A} \right)\,\,.
\eeq

\eq{DICR} can be rewritten as an integral over $z$, namely
\begin{eqnarray}
\sigma_{dipole}\,\,& =& 
\int\,d^2 b_t \,2
N(r,y,b_t)\,\,\,=\,\,2 \pi \,\{\int^{b^2_0}_0 d
b^2_t\,\,N_{\leftarrow}\,\,+\,\,\int^{\infty}_{b^2_0} \,\,d
b^2_t\,\,N_{\rightarrow}\}\,\,; \label{CRZINT1}\\
&=&\,\,\pi\,R^2_A\,\int^{4 \baralpha\,y \,-\,r\,+ \,2\ln\frac{A
R^2_N}{R^2_A}}_0 \,d\,z \,\left( \,1
\,\,-\,\,e^{-\,\zeta(z)}\,\right)\,\,+\,\,2\,\baralpha
\,e\,R^2_A\,\,.\label{CRZINT2}
\end{eqnarray}

It is interesting to notice that $\sigma^A_{dipole}$ for  nucleus target
manifests itself a remarkable scaling in the kinematic region to the left
of the critical line.
Indeed, from \eq{CRZINT2} one can conclude that
\beq \label{SCALING}
\frac{\sigma^A_{dipole}}{\pi  R^2_A}\,\,
=\,\,\frac{\sigma^N_{dipole}\left(\,
x_B, \frac{Q^2\,(R^2_A)^2}{A^2\,(R^2_N)^2} \,\right)}{\pi R^2_N}\,\,.
\eeq

Fig.5 shows the $Q^2$ behaviour of the ratio
$\frac{\sigma^A_{dipole}}{\pi R^2_A}$ at $x=10^{-4}$ with $y =
\ln(0.01/x)$
for different nuclei. The full line describes the  nucleon target and
other lines correspond to nuclei with A= 40 (Ca), 150, 296 (gold) going
from bottom to
the top. One can see that the cross sections are quite larger than the
geometrical estimates. 

\begin{figure}[htbp]
\begin{center}
  \epsfig{file=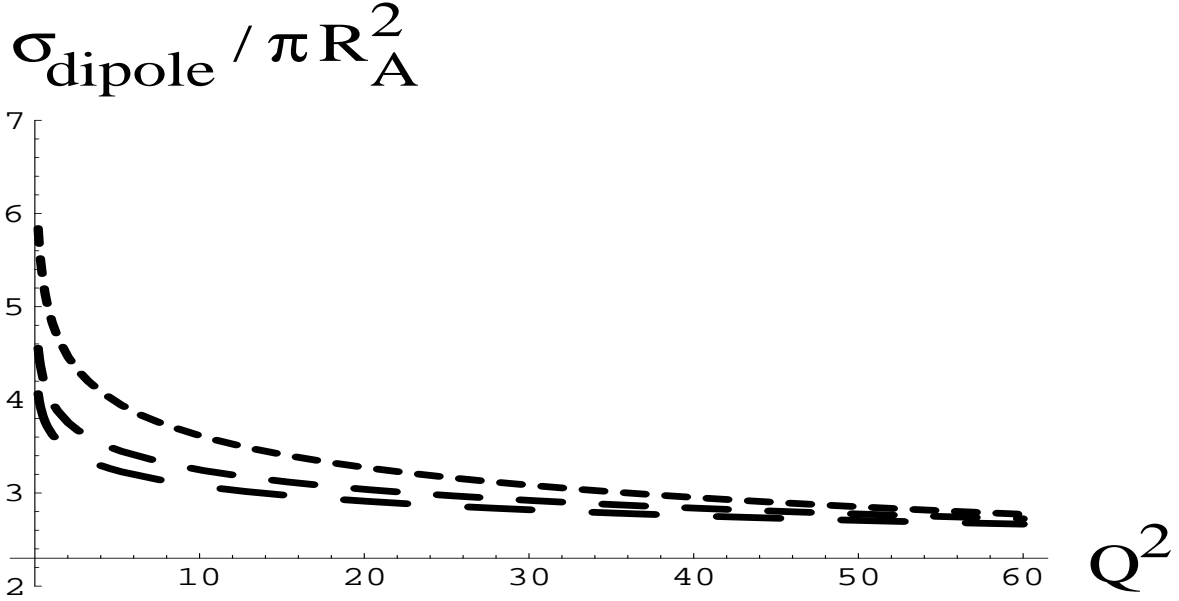,width=160mm, height=80mm}
  \caption[]{\it The ratio of the dipole cross section to the geometrical 
estimates $ \frac{\sigma^A_{dipole}}{\pi R^2_A}$  for  low $x = 10^{-4}$
( $y = \ln(0.01/x)$ ) as a function of $Q^2$ for different nuclei.
All other curves correspond to
A=40, 150, 300  going from the bottom to the top; $\alpha_S = 0.25, Q^2_0 
= 1\,GeV^2$ . }
 \end{center}
\label{fig5}
\end{figure}

It should be stressed that \eq{B0} is an artifact of the oversimplified
 Gaussian parameterization for the nucleus profile function ( see
\eq{PROFFUN}. In more realistic approach to the nucleus profile function
( for example the Wood-Saxon one \cite{WS} )  instead of $\exp( - b^2_t
/R^2_A)$ in \eq{PROFFUN} we have a function which is equal to 1 for $b_t
\leq R_A$ and falls down as $\exp( - b_t/h)$, where $h$ does not depend on
$A$,  at $ b_t > R_A$. Such function
gives $b_0(y,r)$ 
\beq \label{B01}
b_0(y,r)\,\,=\,\,R_A \,\,+ \,\,\frac{h}{2}\,\,\left(\,4\,\baralpha \,y -
r \,+\,2\ln\frac{A R^2_N}{R^2_A}\,\right)\,\,,
\eeq
instead of \eq{B0}\footnote{We are very grateful to Al Mueller for
pointing us a model feature of \eq{B0}.}.

Therefore, we expect the following asymptotic behaviour for
$\sigma_{dipole}$ and $xG(x,Q^2)$  
\begin{eqnarray}
\sigma_{dipole} & \rightarrow & 2 \,\pi \left( R \,\,+\,\,\frac{h}{2}
\,\{\,4\,\baralpha \,y\,-\,r\,+ 2\ln\frac{A
R^2_N}{R^2_A}\,\}\,\right)^2;\label{B03}\\
xG(x,Q^2) & \rightarrow & Q^2 \,\left( R
\,\,+\,\,\frac{h}{2}\,\{\,4\,\baralpha \,y\,-\,r\,+
2\ln\frac{AR^2_N}{R^2_A}\,\}\,\right)^2.\label{B04}
\end{eqnarray}

\section{Summary}

In this paper we found the solution to the evolution equation for high
parton density QCD ( see \eq{GLRINT} ). This solution is given by two
equations: \eq{EVGFSOL} to the right of the critical line ( see Fig.1) and
\eq{ZETA} to the left of the critical line described by \eq{CRLINEFIN}.

This solution gives the colour dipole density ($\rho = x G_A(x,
\frac{1}{{\mathbf{x^2}}})/\pi R^2_A$ ) which is small to the right
of the critical line , reaches the value of the order of unity on the
critical line and increases up to the value $\rho \,\approx\,\,1/\as $ at
very small values of $x$.

Such a behaviour of $\rho$ reveals itself in the following properties of
the gluon structure function:
\begin{enumerate}
\item\,\,\,To the right of the critical line $xG$ is given by the DGLAP
evolution equations;

\item\,\,\,On the critical line $ xG(x,Q^2_{cr}(x))\,=\,Q^2_{cr}(x)R^2_A$
(see \eq{XGCRR} ) where $,Q^2_{cr}(x)$ is defined by \eq{Q0X};

\item\,\,\, To the left of the critical line $xG(x,Q^2) = Q^2 \,R^2_A (
4 \baralpha y - r + 2 \ln\frac{A R^2_N}{R^2_A} )$.
\end{enumerate}

As far as A-dependence is concerned, in the kinematic region to the right
of the critical line we have $xG_A \,\propto\,A xG_N$, while on the
critical line $xG_A(x,Q^2_{cr}(x))\,\propto\,A^{\frac{4}{3}}$ and only to
the left of the critical line we have
$xG_A\,\,\propto\,\,Q^2 R^2_A\,\,\propto\,\,A^{\frac{2}{3}}$.

Our solution reproduces the saturation  of the  gluon density
but due to sufficiently mild dependence on the impact parameter the
saturation leads to the dipole-target total cross section proportional to
$\ln(1/x)$ in the region of the extremely low $x$ (see \eq{DICR1} for the
exact answer).

Both the exact solution near the saturation limit (see \eq{ZETAN} ) and
the proportionality to $xG \,\propto\,Q^2R^2_A$ ( see \eq{GAN} ) are the
manifestation of the fact that nonlinear equation leads to the parton
distribution which peaks at the size
${\mathbf{x^2}}\,\approx\,1/Q^2_{cr}(x)$. In other words, in the region of
low $x$ the parton distribution has a definite mean transverse momentum,
namely, $< p^2_t > = Q^2_{cr}(x)$. Therefore, this solution supports
the more intuitive  approach that has been discussed before for the parton
distribution at
low $x$ \cite{LERY} \cite{MU99}.

We want also to  draw your attention to the fact that in the region of
small $x$  our solution leads to a new scaling between DIS with nucleon
and DIS with nucleus  given by \eq{SCALING}. This equation allows us to
calculate how nucleus DIS approaches $A^{\frac{2}{3}}$ dependence in the
region of low $x$.

We hope, that our solution will stimulate a more detailed study of the
properties of the high density parton system and, in particular, it will 
lead to more quantitive development of  ideas suggested in
Ref\cite{MU99}.

\begin{appendix}
\section{Appendix}
\setcounter{equation}{0}

In this appendix we calculate the phase $\psi(\omega,r,t,b_t)$ in
\eq{SADDLE} at $t\,\ll\,1$.
Using \eq{F}, we can find $p$ for $\as \,\ll\,1$ and $ t
\,\rightarrow\,0$
\beq \label{PVST}
p\,=\,1\,+\,t\,\,-\,\,\frac{\baralpha}{\omega}\,(1+2t)\,\,,
\eeq
which leads to
\begin{eqnarray}
\psi(\omega,r,t,b_t)&+&t ( \eta_A  + \tilde{\eta} ) =
\frac{\baralpha\gamma}{\omega}\,e^{ -\eta_A  } 
(p-1)+pr+\omega y+t ( \eta_A  + \tilde{\eta} )\nonumber  \\
  &\approx& \omega y +r -\frac{\baralpha}{\omega}r-
  tr\left(\frac{\baralpha\gamma}{\omega(-r)}\,e^{- \eta_A  
}\,\,-\,\,1\,\,+\,\,
  \frac{2\baralpha}{\omega}+\frac{1}{r} ( \eta_A  + \tilde{\eta}
)\,\right)\,\,.\label{PSISADDLE}
\end{eqnarray}
Integration over $t$ gives a $\delta$ - function which defines the value
of $\omega = \omega_{cr}$.
\beq \label{OMEGACR}
\omega_{cr}\,=\,\frac{2\baralpha+\frac{\baralpha\gamma}{r}\,e^{ -\eta_A}}
{1-\frac{1}{r} (\eta_A  + \tilde{\eta})}\,\,=\,\,\frac{2
\,\baralpha\,\,\varsigma}{1-\frac{1}{r} (\eta_A  + \tilde{\eta})}\,\,,
\eeq
where $\varsigma\,=\,1 +\frac{\gamma}{2 \,r}e^{- \eta_A}$.

We can make the integration over $\omega = \omega_{cr} + 
\Delta $ , considering $\frac{\Delta}{\omega_{cr}}\,\,\ll\,\,1$. 
\begin{eqnarray}
&&\psi(\omega,r,t,b_t) \,\nonumber\\
&\approx&
- r + \frac{\baralpha r}{\omega_{cr}} +\omega_{cr}y+\Delta\left(
y - \frac{\baralpha r}{\omega_{cr}^2} - \frac{2\baralpha
tr}{\omega_{cr}^2}
\right)\nonumber\\
&=& -  r+ \frac{\baralpha r}{\omega_{cr}} +\omega_{cr}y+
\Delta\frac{2\baralpha (r)}{\omega_{cr}^2}
\left(\underbrace{
(y - \frac{\baralpha r}{\omega_{cr}^2})
\frac{\omega_{cr}^2}{2\baralpha(r)\,\varsigma}}_{t_0}
-t\right) \,\,.\label{OMSAD}
\end{eqnarray}
Integration over $\Delta$ gives
$\delta\left[\frac{2\baralpha(-r)}{\omega_{cr}^2}(t-t_0)\right]$ which
yields
\begin{equation} \label{SOLCRLINE}
g_A(y,\xi=\eta_A,r,b_t)\,=\,\frac{\omega^2_{cr}}{2 \pi\, 
2\baralpha (r) \,t_0\,\varsigma}\,\frac{ \as \pi^2 A}{4 \pi \,R^2_A}
\,e^{-
\frac{b^2_t}{R^2_A}}\,
e^{\omega_{cr}y - r+ \frac{\baralpha r}{\omega_{cr}}} \,\,.
\end{equation}
\end{appendix}

\end{document}